\documentclass{mn2e}

\usepackage{times}
\usepackage{graphicx}

\begin{document}
\journal{astro-ph/yymmnnn}

\title[Planck and spectral index running]{When can the Planck
  satellite measure spectral index running?} 
\author[C. Pahud et al.]
{C\'edric Pahud, Andrew R. Liddle, Pia Mukherjee, and David Parkinson\\ 
Astronomy Centre, University of Sussex, Brighton BN1 9QH, United
Kingdom}
\maketitle
\begin{abstract}
We use model selection forecasting to assess the ability of the Planck
satellite to make a positive detection of spectral index running. We
simulate Planck data for a range of assumed cosmological parameter
values, and carry out a three-way Bayesian model comparison of a
Harrison--Zel'dovich model, a power-law model, and a model including
running. We find that Planck will be able to strongly support
running only if its true value satisfies $|dn/d\ln k| > 0.02$.
\end{abstract}

\section{Introduction}

Results from the Wilkinson Microwave Anisotropy Probe (WMAP),
especially the first-year data \cite{wmap1} and to some extent the
three-year data \cite{wmap3}, have placed a focus on possible running
of the spectral index of density perturbations (see e.g.~Lidsey \&
Tavakol 2003; Kawasaki, Yamaguchi \& Yokoyama 2003; Chung, Shiu \&
Trodden 2003; Bastero-Gil, Freese \& Mersini-Houghton 2003; Chen et
al.~2004; Covi et al.~2004; Ashoorioon, Hovdebo \& Mann 2005;
Ballesteros, Casas \& Espinosa 2006; Cline \& Hoi 2006; Cort\^es \&
Liddle 2006; Easther \& Peiris 2006). It is certainly premature to
draw any strong conclusions based on existing evidence, especially as
it remains controversial whether current data even support power-law
models over the Harrison--Zel'dovich (HZ) model, but it is timely to
investigate the extent to which the upcoming Planck satellite may
resolve the situation.

As we have stressed in several recent papers (e.g.~Mukherjee,
Parkinson \& Liddle 2006a; Parkinson, Mukherjee \& Liddle 2006;
Liddle, Mukherjee \& Parkinson 2006a), the appropriate statistical tool
for assessing the need to introduce new parameters is \emph{model
selection} \cite{Jeff,mackay,gregory}. Model selection assigns
probabilities to \emph{sets} of parameters, i.e.~models, in addition
to the usual probability distributions for parameter values within
each model. For example, Bayesian model selection applied to data
compilations including WMAP3 shows that the case for including even
just the spectral index $n_{{\rm S}}$ as a variable fit parameter is
inconclusive \cite{PML}.

In a recent paper \cite{PLMP}, we used model selection forecasting
tools to assess the ability of the Planck satellite to distinguish
between the Harrison--Zel'dovich model with $n_{{\rm S}} = 1$ and a
model with varying spectral index, VARY$n$. The outcome naturally
depends on the assumed true value of $n_{{\rm S}}$, which we call the
\emph{fiducial value}, and we found that Planck can strongly favour
the latter model only if the true value of $n_{{\rm S}}$ lies outside
the range $[0.986,1.014]$. In making that comparison, we assumed that
the true spectrum could be described by a power-law.

In this paper, we extend that analysis to include the possibility of
spectral index running, given by \mbox{$\alpha \equiv dn/d\ln k$}. This
adds an extra model, VARY$n\alpha$, to the model set. This means that
we are carrying out a three-way model comparison, within the
two-dimensional space defined by the fiducial values of $n_{{\rm S}}$
and $\alpha$. Ideally we would also have included tensor perturbations
in this analysis in order to fully represent the usual inflationary
predictions (e.g.~Liddle \& Lyth 2000), but the present analysis is at
the limits of current computer power, having required many months of
multi-processor time.

\section{Model selection forecasts for models with running}

\subsection{Model selection forecasting}

Our approach exactly follows our earlier paper \cite{PLMP}, and so we
provide only the briefest of summaries here and refer to that paper
and references therein for details. Model selection forecasting was
first introduced by Trotta (2007b), whose Predictive Posterior Odds Distribution (PPOD) forecasting determined
the probability of different model selection outcomes of future
experiments based on present knowledge. An alternative approach, which
delineates regions of parameter space where different model selection
verdicts are expected, was introduced in Mukherjee et al.~(2006b); a
combination of the methods was used in Pahud et al.~(2006), and also
subsequently in Trotta (2006), Liddle et al.~(2006b) and Trotta
(2007a).  As used here, data are simulated for the different possible
true values of the parameters of interest, known as fiducial values,
and a model comparison analysis carried out at each point.

Although not required, in typical examples a simple model will be
nested within a more complex one, e.g.~the HZ model is the special
case of VARY$n$ with $n_{{\rm S}} = 1$. If the (assumed) true model is
the nested one, the model comparison will favour that model, and one
may ask how strongly. If instead the true model is the more complex
one, one can ask how far from the simple model the true values would
have to be, in order that a given experiment can overcome statistical
uncertainty and deliver a strong or decisive verdict in favour of the
complex model. These two notions can be used to define model selection
Figures-of-Merit, assessing the abilities of competing experiments
\cite{MPCLK}.

In our work, we use the Bayesian evidence $E$ as the model selection
statistic. Like any model selection statistic, it creates a tension
between goodness of fit to the data and the complexity of the
model. It represents a full implementation of Bayesian inference,
being the probability of the data given the model (i.e.~the model
likelihood). It updates the prior model probability to the posterior
model probability. Computations are carried out using the nested
sampling algorithm \cite{Skilling}, using our code
\textsc{cosmonest}\footnote{Available at http://cosmonest.org.}
\cite{MPL,PML}. Computing the evidence accurately is significantly
more challenging than computing the posterior probability
distribution, and so the calculations are computationally
time-consuming. 

Our assumption is that there are three models of interest in fitting
future Planck data. These are the Harrison--Zel'dovich model, a
power-law model where $n_{{\rm S}}$ is fit from data, and a model
where both $n_{{\rm S}}$ and $\alpha$ are varied. We denote these
models HZ, VARY$n$, and VARY$n\alpha$ respectively, and also indicate
them by use of subscripts 0, 1, and 2 respectively.

In the presence of running, the spectral index is defined in the usual
way by
\begin{equation}
n_{\rm{S}}(k)=n_{\rm{S}}(k_0)+\alpha \ln  \frac{k}{k_0}\,.
\label{running}
\end{equation}
The pivot scale $k_0=0.05 \, {\rm Mpc}^{-1}$ corresponds to a scale
well constrained by existing data. When running is included, $n_{\rm
S}$ is always specified at this scale, and throughout we assume the
running is constant.  As in Pahud et al.~(2006), the prior range for
$n_{{\rm S}}$ is taken to be $0.8 < n_{{\rm S}} < 1.2$, representing a
reasonable range allowed by slow-roll inflation models (see
e.g.~Liddle \& Lyth 2000).

We take the prior on $\alpha$ to be $-0.1 < \alpha < 0.1$. This is
somewhat arbitrary. Slow-roll inflation models would tend to suggest a
much smaller value \cite{KT}, but there is no point in restricting the
analysis to values smaller than Planck can measure, as one will simply
conclude that Planck is unable to make the measurement. Accordingly,
our range is loosely motivated by present observational knowledge,
corresponding to models with unexpectedly large running.  The
comparison between two models does have some prior dependence on the
extra parameter(s). If one prior is widened in regions where the
likelihood is negligible, then the evidence changes proportional
to the prior volume, so for instance a doubling of the prior range
will only reduce $\ln E$ by $\ln 2 = 0.69$.

In running CosmoNest, the algorithm parameters used were $N = 300$
live points and an enlargement factor of 1.8 for HZ, 1.9 for VARY$n$,
and 2.0 for VARY$n\alpha$. The tolerance parameter was set to 20
(rather than 0.5 as in our previous analysis) in order to improve the
speed of the simulations. This is sufficient to give answers to good
accuracy as indicated by the uncertainties obtained. Four independent
evidence evaluations were done for each calculation, to obtain the
mean and its standard error.

We then compare our models in pairs by considering the Bayes factor,
defined as the ratio of evidences between two models, written $B_{ij}
= E(M_i)/E(M_j)$, for $i,j=0,1,2$ ($i\not= j$), where $M_i$ and $M_j$
indicate the two models under assumption. By plotting the Bayes factor
using datasets generated as a function of the two parameters of
interest, one uncovers the regions of the two-dimensional fiducial
parameter space in which the Planck satellite would be able to
decisively select between the two models, and also those regions where
the comparison would be inconclusive.

In order to assess the significance of any difference in evidence
between two models, a useful guide is given by the Jeffreys' scale
\cite{Jeff}. Labelling as $M_i$ the model with the higher evidence, it
rates \mbox{$\ln B_{ij} < 1$} as `not worth more than a bare mention',
$1<\ln B_{ij} < 2.5$ as `substantial', $2.5< \ln B_{ij} < 5$ `strong'
to `very strong', and $5<\ln B_{ij}$ as `decisive'.

\subsection{Simulating Planck data}

We simulate Planck data exactly as described in Pahud et
al.~(2006). Having determined the fiducial model power spectra, we
simulate temperature power spectrum data for the three most sensitive
High Frequency Instrument (HFI) channels and the polarization signal for only one of these
channels, modelling instrument noise using current detector
specifications. The simulations are somewhat simplistic, as
computational limitations prevent a more detailed treatment that might
include residuals from foreground subtraction and $1/f$ noise. However
they should provide a good characterization of the Planck data for our
purposes. Simulations are carried out for various values of the
spectral index and its running, and the other parameters are those of
the usual $\Lambda$CDM model in a flat spatial geometry.

\begin{figure}
\centering
\includegraphics[width=7.5cm]{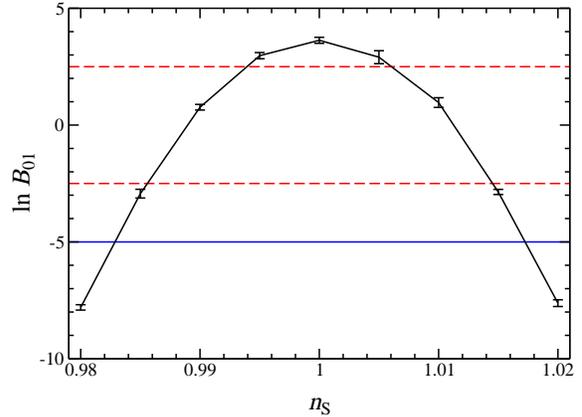}
\caption{\label{f:plot_lnB_01_1D} The logarithm of the Bayes factor, $\ln
B_{01}$, as a function of the fiducial value of $n_{{\rm S}}$. The
horizontal lines indicate where the comparison becomes `strong'
(dashed) and `decisive' (solid) on the Jeffreys' scale.}
\end{figure}

In simulating the data, we are primarily interested in the dependence
on the key parameters of interest, $n_{{\rm S}}$ and $\alpha$, and
different data simulations are carried out for a grid of values in
that plane. The other cosmological parameters are given fixed fiducial
values as in Pahud et al.~(2006), namely the baryon physical density
$\Omega_{\rm b}h^2 = 0.024$, the cold dark matter physical density
$\Omega_{\rm c} h^2 = 0.103$, the sound horizon $\Theta = 1.047$, the
optical depth $\tau = 0.14$, and the density perturbation amplitude
normalization $A_{\rm S}= 2.3 \times 10^{-9}$. The corresponding value
of the Hubble parameter is $h=0.78$. The model selection verdict
should have negligible dependence on these fiducial values. Note that
all parameters, including these, are varied in computing the evidences
of the models; it is only in defining the fiducial models for data
simulation that these parameters are fixed. The prior ranges used for
these parameters are as in Pahud et al.~(2006): $0.018 \leq
\Omega_{\rm b}h^2 \leq 0.032$, $0.04 \leq \Omega_{\rm c} h^2 \leq
0.16$, $0.98 \leq \Theta \leq 1.1$, $0 \leq \tau \leq 0.5$, and $2.6
\leq \ln(A_{\rm S} \times 10^{10}) \leq 4.2$.

\begin{figure*}
\centering
\includegraphics[width=5.4cm]{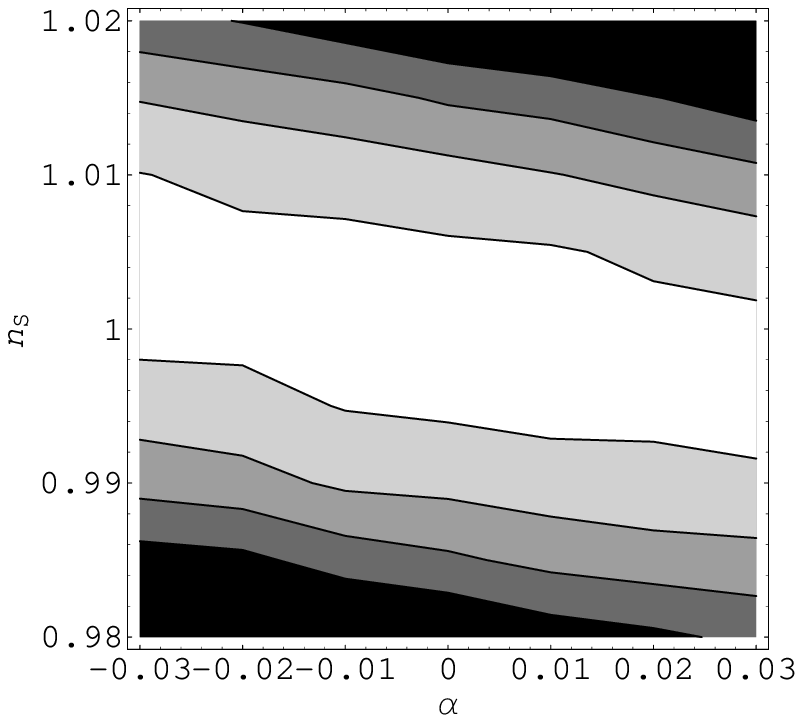}
\hspace*{0.2cm}
\includegraphics[width=5.4cm]{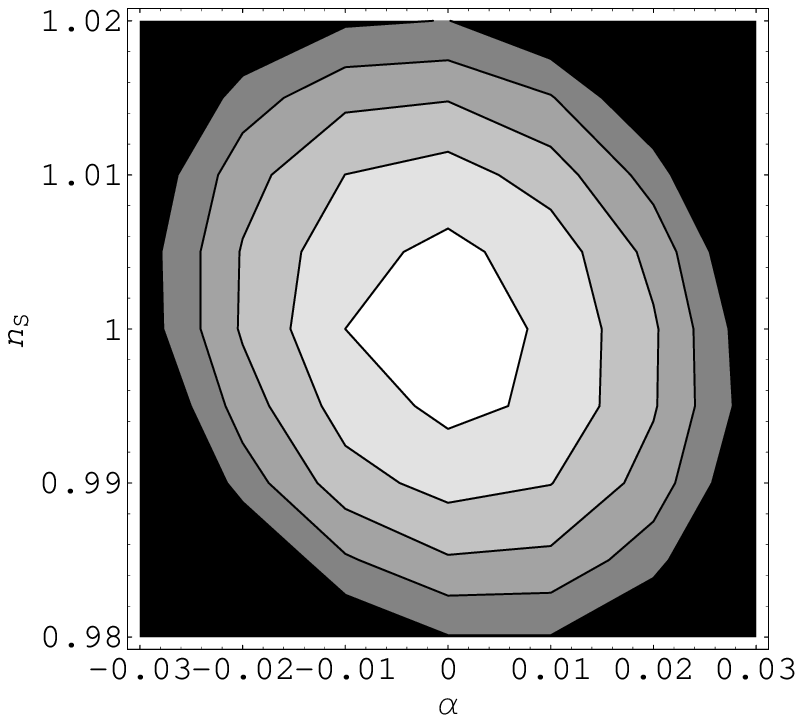}
\hspace*{0.2cm}
\includegraphics[width=5.4cm]{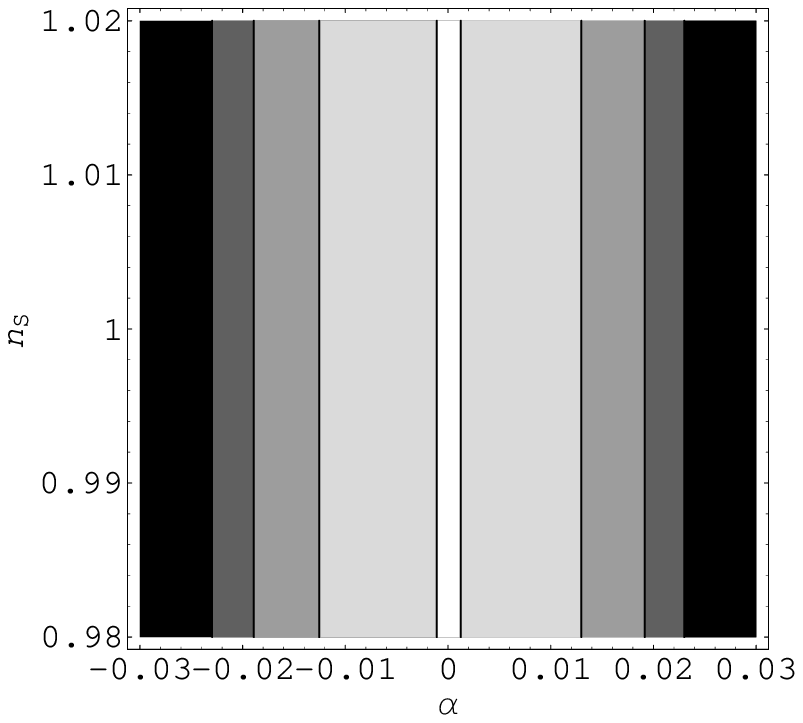}
\caption{\label{f:plot_lnB_01} The logarithm of the Bayes factors,
$\ln B_{01}$ in the left panel, $\ln B_{02}$ in the centre, and $\ln
B_{12}$ in the right, as a function of the fiducial values of $n_{{\rm
S}}$ and $\alpha$. The contour lines represent different steps in the
Jeffreys' scale. From the plot centres, the levels are 2.5, 0, -2.5,
-5 in the left and right panels, with the centre panel contours
starting at 5.}
\end{figure*}

\section{Results}

We begin by showing in Fig.~\ref{f:plot_lnB_01_1D} the main result
obtained in Pahud et al.~(2006). In that analysis, running was not
included and so the fiducial $\alpha$ is zero. At $n_{{\rm S}} = 1$,
corresponding to HZ being the true model, the HZ model is strongly
preferred with $\ln B_{01}= 3.6 \pm 0.1$. It has a higher evidence
since it can fit the data just as well as VARY$n$ and has one less
parameter. Once $n_{{\rm S}}$ is far enough away from 1, the HZ fit
becomes very poor and the Bayes factor plummets. The speed with which
this happens indicates the strength of the experiment. The VARY$n$
model becomes strongly favoured only once $n_{{\rm S}} < 0.986$ or
$n_{{\rm S}} > 1.014$; if the true value lies within that range even
the Planck satellite will give inconclusive results.

Figure~\ref{f:plot_lnB_01} shows the extension of our results into the
\mbox{$\alpha$--$n_{{\rm S}}$} plane, now showing the three-way model
comparison. The left plot still shows the comparison between HZ and
VARY$n$, though neither is the true model except at $\alpha = 0$
(Fig.~\ref{f:plot_lnB_01_1D} is the cross-section of this plot at
$\alpha=0$).  The plot is not surprising in the sense that the
logarithm of the Bayes factor is roughly independent of $\alpha$. The
models HZ and VARY$n$ are just as bad at describing a non-zero
running. However, a slight tilt of the contours appears when $\alpha$
goes away from zero. This indicates that a positive (resp.\ negative)
running can be balanced by a scalar index smaller (resp.\ bigger) than
$1$, accordingly to equation (\ref{running}). This can benefit HZ or
VARY$n$, depending whether it helps or hinders the HZ model to fit the
data. In fact the effect just reflects that the scale $k_0$ is not
quite at the statistical centre of the data, so that the determination
of $n_{{\rm S}}$ and $\alpha$ has some correlations, and could be
removed by judicious choice of the `pivot' scale (Cort\^es, Liddle \&
Mukherjee 2007).

The centre panel now introduces a comparison of HZ with VARY$n\alpha$,
which is the true model in most of the parameter plane.  At
[$\alpha,n_{{\rm S}}$]=[0,1], the HZ model is decisively preferred
with $\ln B_{02}= 6.3 \pm 0.1$. Its higher evidence arises since it
can fit the data just as well as VARY$n\alpha$, but has two less
parameters. Once the fiducial point in the two-dimensional space is
far enough away from the centre, the HZ fit becomes very poor and
VARY$n\alpha$ model becomes favoured.

Being the true model, VARY$n\alpha$ can simply adapt its two extra
free parameters to fit the data at every point of the fiducial space
equivalently, thus leading to the same evidence. We have verified this
holds to excellent accuracy in our simulations.  The behaviour of the
Bayes factor should therefore be approximately symmetrical with
respect to $n_{{\rm S}}=1$ and to $\alpha=0$. However, it is clearly
not quite the case, for the same reason as the presence of the tilt in
the left panel. The influence of the correlation between the two
fiducial parameters is greater this time, as it acts on HZ only.

\begin{figure}
\centering
\includegraphics[width=7.5cm]{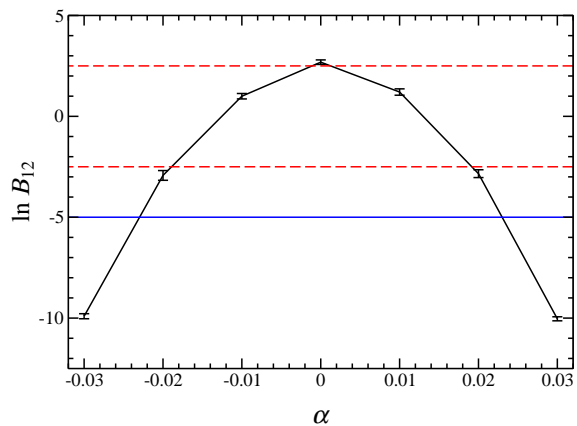}
\caption{\label{f:plot_lnB_12_1D} The logarithm of the Bayes factor,
$\ln B_{12}$, as a function of the fiducial value of $\alpha$. The
horizontal lines indicate where the comparison becomes `strong'
(dashed) and `decisive' (solid) on the Jeffreys' scale.}
\end{figure}

Finally, we need to consider a comparison between the models VARY$n$
and VARY$n\alpha$, which is illustrated in the right panel of
Fig.~\ref{f:plot_lnB_01}. This plot is fully determined by the above
results, as by definition $\ln B_{12}=\ln B_{02}-\ln
B_{01}$. Moreover, for the same reason that the evidence of
VARY$n\alpha$ is independent of both fiducial parameters $n_{{\rm S}}$
and $\alpha$, VARY$n$ turns out to be independent of $n_{{\rm
S}}$. This allows us to restrict our analysis to one dimension only,
shown in Fig.~\ref{f:plot_lnB_12_1D}. At $\alpha=0$ the VARY$n$ model
is strongly preferred over VARY$n\alpha$ as $\ln B_{12}= 2.7 \pm 0.1$,
having one less parameter. The running model becomes strongly favoured
only if the true running satisfies $|\alpha| > 0.02$.

\begin{figure}
\centering
\includegraphics[width=7.5cm]{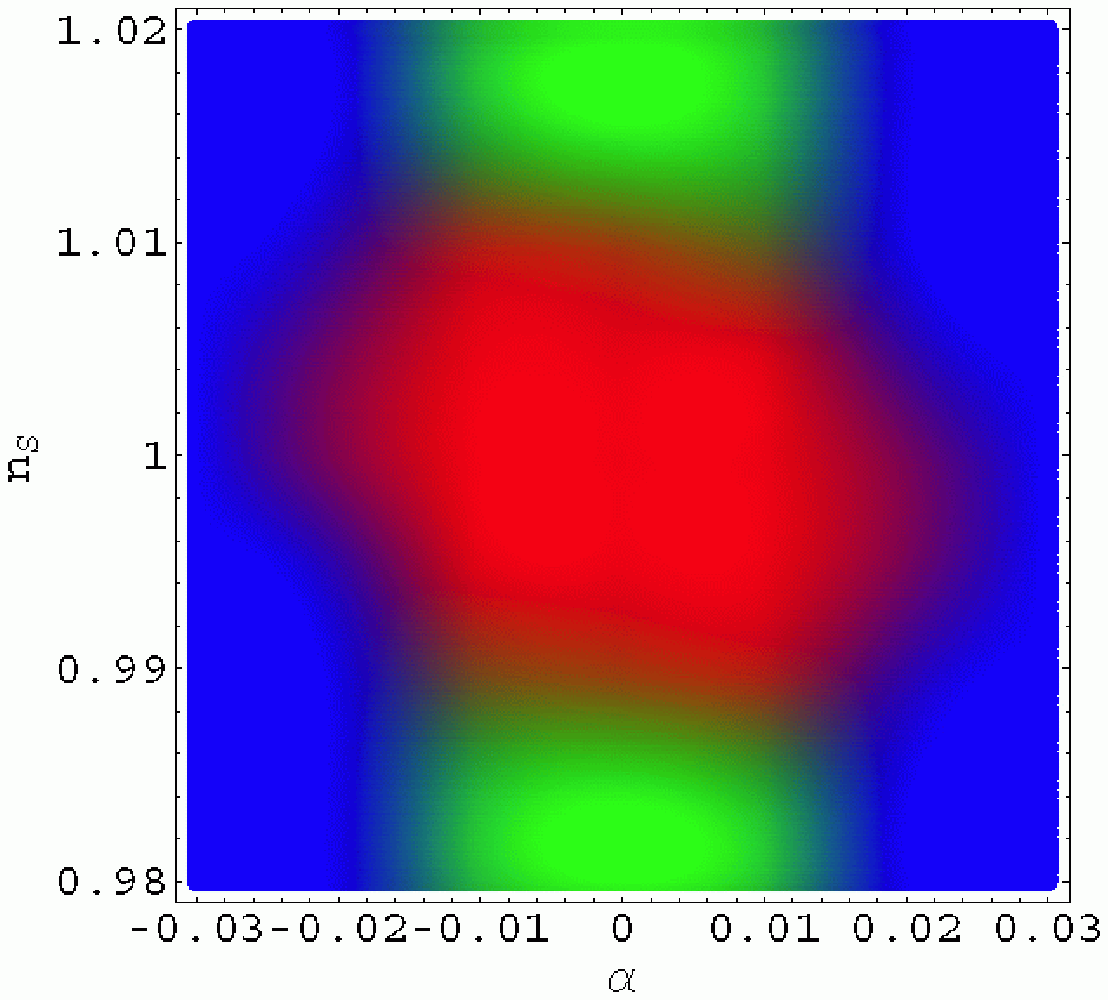}\\
\includegraphics[width=7.5cm]{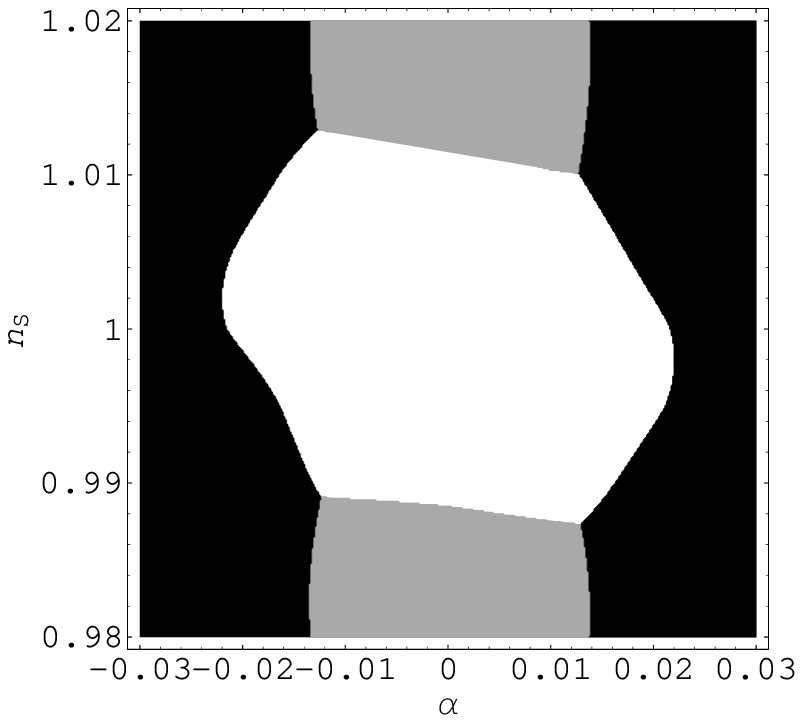}
\caption{\label{f:RGB} Two graphical representations of the three-way
model comparison. The upper panel is a false-colour RGB plot with the
probabilities of HZ, VARY$n$, and VARY$n\alpha$ assigned to the red,
green, and blue channels respectively. The lower panel simply shows
the model which would receive the highest model probability at each
point in the fiducial parameter space, with those three models
allocated white, grey, and black respectively.}
\end{figure}

In Fig.~\ref{f:RGB}, we display the full three-way model comparison in
two different ways. The three-model case is perfectly adapted to
display by false-colour RGB plot, where the intensity of each of the
three red--green--blue colour channels at a given fiducial point is
assigned as the posterior model probability, given by Bayes' theorem,
\begin{equation}
P_i = P(M_i|D) = \frac{P(D|M_i)P(M_i)}{\sum_j P(D|M_j)P(M_j)} \,.
\end{equation}
Here we assume that the prior model probabilities $P(M_i)$ are equal
(an assumption readily varied if required), so the equation simplifies
to
\begin{equation}
P_i = \frac{E(M_i)}{\sum_j E(M_j)} \,.
\end{equation}
That the total probability
sums to one corresponds to fixed total intensity. This is shown in the
upper panel. The region which appears red would lead to the HZ model
being preferred, green the VARY$n$ model, and blue the VARY$n\alpha$
model. Between those, regions which interpolate into secondary colours
share their probability between the different models. There are also
four `vertices' at which all three models have the same
probability. We see that the transitions between the different domains
are rather rapid in terms of the shifting model probabilities.

The lower plot shows a much simpler representation, where regions are
shaded simply according to the dominant model probability in that
region.

These two plots affirm the results already apparent from the earlier
figures; for Planck to be able to demonstrate that $n_{{\rm S}} \neq
1$, the true value will have to be more than 0.01 away from unity
\cite{PLMP}, and for running to be convincingly detected $|\alpha|$
will need to be at least 0.02.

\section{Conclusions}

According to WMAP3 analyses \cite{wmap3}, the running is presently
constrained, at 95\% confidence, to be in the range of approximately
$-0.17 < \alpha < +0.01$. The precise constraints depend on both on
the dataset combination used and the model assumptions made
(e.g.~whether or not to include tensor perturbations), and we have
simply quoted the broadest available. Although the range is highly
skewed to negative values, the special status of $\alpha = 0$, and the
prediction from slow-roll inflation for an $\alpha$ value that current
experiments cannot distinguish from zero, means that from a model
selection point of view $\alpha=0$ should still be regarded as a very
plausible interpretation of the data.

Given this inconclusive position, we have addressed the extent to
which the Planck satellite is likely to resolve the situation, using
model selection tools to compare three models: Harrison--Zel'dovich
(HZ), power-law initial perturbations (VARY$n$), and the running model
(VARY$n\alpha$). The expected outcome depends, of course, on which (if
any) of these models proves to be the correct one.

Supposing first that HZ is the true model, we found in
Pahud et al.~(2006) that VARY$n$ would be strongly, though not
decisively, disfavoured after Planck. The present paper adds the new
information that the running model would be decisively disfavoured in
this circumstance.

Suppose instead that VARY$n$ is true. Then VARY$n$ will be strongly,
but not decisively, preferred over VARY$n\alpha$. However, as shown in
Pahud et al.~(2006), the true value of $n_{{\rm S}}$ has to be
sufficiently far from one in order for VARY$n$ to be favoured over
HZ. Depending on the true parameter values, all three models may
survive application of Planck data.

Finally, suppose VARY$n\alpha$ is true. The alternatives will only be
decisively ruled out provided the true value satisfies $|\alpha| \ga
0.02$, otherwise the outcome will again be indecisive. The conclusion
is that Planck will improve knowledge as compared to WMAP3, by a
factor of around four (our calculations indicate a projected parameter
uncertainty from Planck of about $\pm 0.007$ on $\alpha$, to be
compared with the current $\pm 0.03$ from WMAP3 alone), and thus does
have the capability to convincingly detect running if it is
prominent. However, it does not have the accuracy to probe into the
region where slow-roll inflation models typically lie (Kosowsky \&
Turner 1995).

Our analysis refers to Planck satellite data alone, and, as with
WMAP3, one would expect some further tightening with incorporation of
other datasets probing different length scales.

As with any Bayesian analysis, the results have some dependence on prior
assumptions. For the priors we have chosen on $n_{{\rm S}}$ and
$\alpha$, the data are able to constrain the likelihood well within
them. Consequently, any change in prior ranges that continues to
respect this will just change the evidences according to the change in
volume, an effect one can readily calculate. Bearing in mind that the
Jeffreys' scale is logarithmic, a sizeable change in prior parameter
ranges would be needed to significantly alter the conclusions.

\section*{Acknowledgments}
C.P.\ was supported in part by the Swiss Sunburst Fund, and A.R.L.,
P.M., and D.P.\ by PPARC.  C.P.\ acknowledges the hospitality of
Caltech and Marc Kamionkowski while this work was completed, during a
visit supported by the Royal Astronomical Society and by
Caltech. A.R.L.\ acknowledges the hospitality of the Institute for
Astronomy, University of Hawai`i, while this work was being completed.
We thank Jim Cline for prompting us to look at this issue.

\bsp


\begin{thebibliography}{}
\bibitem[Ashoorioon et al.~2005]{AHM} Ashoorioon A., Hovdebo J. L.,
  Mann R. B., 2005, Nucl. Phys. B, 727, 63, gr-qc/0504135
\bibitem[Ballesteros et al.~2006]{BCE} Ballesteros G., Casas J. A.,
  Espinosa J. R., 2006, JCAP, 0603, 001, hep-ph/0601134
\bibitem[Bastero-Gil et al.~2003]{BFM} Bastero-Gil M., Freese K., 
Mersini-Houghton L., 2003, Phys. Rev. D, 68, 123514,  hep-ph/0306289
\bibitem[Chen et al.~2004]{Cetal04} Chen. C., Feng B., Wang X., Yang
  Z., 2004, Class. Quant. Grav., 21, 3223, astro-ph/0404419
\bibitem[Chung et al.~2003]{CST} Chung D. J. H., Shiu G., Trodden M.,
  2003, Phys. Rev. D, 68, 063501, astro-ph/0305193
\bibitem[Cline \& Hoi 2006]{CH06} Cline J. M., Hoi L., 2006, JCAP,
  0606, 007, astro-ph/0603403
\bibitem[Cort\^es \ & Liddle 2006]{CL06} Cort\^es M., Liddle A. R.,
 2006, Phys. Rev. D, 73, 083523, astro-ph/0603016
\bibitem[Cort\^es et al 2007]{CLP07} Cort\^es M., Liddle A. R.,
  Mukherjee P., 2007, Phys. Rev. D, 75, 083520, astro-ph/0702170
\bibitem[Covi et al.~2004]{CLMO} Covi L., Lyth D. H., Melchiorri A.,
  Odman C. J., 2004, Phys. Rev. D, 70, 123521, astro-ph/0408129
\bibitem[Easther \& Peiris 2006]{EP06} Easther R., Peiris H., 2006,
  JCAP, 0609, 010, astro-ph/0604214
\bibitem[Gregory 2005]{gregory} Gregory P., 2005, \emph{Bayesian
	logical data analysis for the 
	physical sciences}, Cambridge University Press 
\bibitem[Jeffreys 1961]{Jeff} Jeffreys H., 1961, \emph{Theory of
	Probability}, 3rd ed, Oxford University Press 
\bibitem[Kawasaki et al.~2003]{KYY} Kawasaki M., Yamaguchi M.,
  Yokoyama J., 2003, Phys. Rev. D, 68, 023508, hep-ph/0304161
\bibitem[Kosowsky \& Turner 1995]{KT} Kosowsky A., Turner M. S., 1995,
  Phys. Rev. D, 52, 1739, astro-ph/9504071
\bibitem[Liddle \& Lyth 2000]{LL} Liddle A. R., Lyth D. H., 2000, {\em
	Cosmological inflation 	and large-scale structure}, Cambridge
	University Press 
\bibitem[Liddle et al.~2006a]{LMP} Liddle A. R., Mukherjee P.,
  Parkinson D., 2006a, A\&G, 47, 4.30, astro-ph/0608184
\bibitem[Liddle et al.~2006b]{LMPW} Liddle A. R., Mukherjee P.,
  Parkinson D., Wang Y., 2006b, Phys. Rev. D, 74, 123506,
  astro-ph/0610126 
\bibitem[Lidsey \& Tavakol 2003]{running} Lidsey J. E., Tavakol R.,
  2003, Phys. Lett. B, 575, 157, astro-ph/0304113
\bibitem[MacKay 2003]{mackay} MacKay D. J. C., 2003, \emph{Information
	theory, inference and learning algorithms}, Cambridge
	University Press 
\bibitem[Mukherjee et al.~2006a]{MPL} Mukherjee P., Parkinson D.,
	Liddle A. R., 2006a, ApJL, 638, L51, astro-ph/0508461
\bibitem[Mukherjee et al.~2006b]{MPCLK} Mukherjee P., Parkinson D.,
  Corasaniti P. S., Liddle A. R., Kunz M., 2006b, MNRAS, 369, 1725,
  astro-ph/0512484
\bibitem[Pahud et al.~2006]{PLMP} Pahud C., Liddle A. R., Mukherjee
	P., Parkinson D., 2006,	Phys. Rev. D, 73, 123524,
	astro-ph/0605004 
\bibitem[Parkinson et al.~2006]{PML} Parkinson D., Mukherjee P.,
  Liddle A. R., 2006, Phys. Rev. D, 73, 123523, astro-ph/0605003
\bibitem[Skilling 2006]{Skilling} Skilling J., 2006, Bayesian Anal.,
  {\bf 1}, 833
\bibitem[Spergel et al.~2003]{wmap1} Spergel D. N. et al.~(the WMAP
  Team), 2003, ApJS, 148, 175, astro-ph/0302209
\bibitem[Spergel et al.~2007]{wmap3} Spergel D. N. et al.~(the WMAP
	Team), 2007, ApJS, 170, 377, astro-ph/0603449
\bibitem[Trotta 2006]{Trottaconf} Trotta R., 2006, in proceedings of
  ``Cosmology, galaxy formation and astroparticle physics on the
  pathway to the SKA'', astro-ph/0607496
\bibitem[Trotta 2007a]{Trotta} Trotta R., 2007a, MNRAS, 378, 72, 
astro-ph/0504022
\bibitem[Trotta 2007b]{Trotta2} Trotta R., 2007b, MNRAS, 378, 819,
astro-ph/0703063
\end{thebibliography}
\end{document}